\documentstyle[twocolumn,aps,pra,epsfig]{revtex}

\begin{document}

\wideabs{

\title{Dissipative dynamics of vortex arrays in anisotropic traps}

\author{O.N. Zhuravlev\protect\( ^{1}\protect \), A.E. Muryshev\protect\( ^{1}\protect \)
and P.O. Fedichev\protect\( ^{1,2}\protect \)}

\address{\protect\( ^{1)}\protect \)Russian Research Center ``Kurchatov Institute'',
Kurchatov Square, 123182, Moscow, Russia}

\address{\protect\( ^{2)}\protect \)Institut f. Theoretische Physik, Universitat Innsbruck,
Technikerstr. 25/2, A6020, Innsbruck, Austria}

\maketitle
\begin{abstract}
We discuss the dissipative dynamics of vortex arrays in trapped Bose-condensed
gases and analyze the lifetime of the vortices as a function of trap anisotropy
and the temperature. In particular, we distinguish the two regimes of the dissipative
dynamics, depending on the relative strength of the mutual friction between
the vortices and the thermal component, and the friction of the thermal particles
on the trap anisotropy. We study the effects of heating of the thermal cloud
by the escaping vortices on the dynamics of the system.
\end{abstract}
}

\section{Introduction}

Recent experimental evidence \cite{vortobserv:cornell,vortobserv:dalibard,vortobserv:foot}
for vortices in weakly interacting Bose-condensed gases stimulated a number
of publications on structure of a rotating state of superfluids (see \cite{Pitaevskii:review}
for a review). Latest state of the art experimental technology makes possible
creation and study of sufficiently large vortex arrays \cite{dalibard:varray},
as well as non-destructive measurements of time evolution in a system with a
single vortex \cite{cornell:vortexdyn}. Such possibilities allow, in principle,
experimentally clarify the nature of interaction of vortices with thermal excitations.
This issue plays an important role for understanding of vortex dynamic in numerous
related systems, ranging from superconductors to superfluid cores of neutron
stars (see \cite{Sonin} for a review), and still remains a controversial issue
in the literature \cite{vinen:puzzle}.

In our recent papers we investigated the dissipative dynamics of a single vortex
\cite{fedichev:vortexdyn} and a large vortex array \cite{fedichev:varray1}
in a static (non-rotating) trap at finite temperatures. By this we assumed a
very fast relaxation of the thermal component towards its static configuration
due to, say, the anisotropy of the trap. Since actual experiments are performed
in traps, which can be made nearly perfect, the rotation of the thermal component
can persist for a very long time. Moreover, in the case of a large vortex array,
the decay of vortices reaching the border of the condensate may be accompanied
by a substantial heating of the thermal cloud, which, in turn, affects the strength
of the friction forces acting on the remaining vortices. Thus, in order to understand
an experiment in realistic conditions and use its results for a fine test of
vortex dynamics theories, one has to have a solid description of dissipative
dynamics including various temperature effects as well as the role of the trap
anisotropy.

In this Letter we develop a simple, but yet a feature rich, model of dissipative
dynamics of large vortex arrays in realistic anisotropic traps at finite temperatures.
We analyze the relaxation kinetics of a system consisting of the rotating thermal
component interacting with a condensate the vortices. We show that the trap
anisotropy, or any other mechanism of angular momentum dissipation, is absolutely
crucial for the onset of the dissipative dynamics. On the basis of our analysis
we distinguish the two regimes of the evolution. The first one corresponds to
the previously studied static case \cite{fedichev:varray1}, i.e. to the situation
when the trap is sufficiently anisotropic and the rotation of the thermal cloud
quickly ceases. In this case the vortex array decays in a non-exponential fashion
(a power law for the number of vortices at sufficiently long times). Remarkably,
in this limit the lifetime of vortices does not depend on the parameters of
anisotropy. In the opposite case, i.e. when the trap is nearly perfect, the
friction of the vortices on the thermal cloud is stronger than the the friction
of the thermal cloud on the trap asymmetry. This makes the vortices and the
cloud stick to each other: the condensate rotates with the same angular velocity
as the thermal particles and the angular velocity of the both components decreases
exponentially with a time constant explicitly depending on the parameters of
the trap asymmetry. 

We discuss the heating of the thermal component due to the vortex array decay.
The developed model, together with certain simplifications, allows us to obtain
a simple analytical description of dissipative dynamics and compare the results
with the experimental data \cite{dalibard:varray}. We believe that the presented
model can be used for indirect study of vortex arrays dynamics based on the
thermal component thermometry (proposed earlier in \cite{fedichev:varray1}).

\section{The description of the model}

We consider an evolving vortex array in an anisotropic trap characterized by
the radial and the longitudinal frequencies \( \omega _{\rho } \) and \( \omega _{z} \).
The equilibrium vortex configuration in a rotating trap (in the experiment the
\cite{dalibard:varray} this corresponds to the situation when the stirring
beam is on) is considered in \cite{LL:volIX}. Further on we assume that the
initial angular velocity of the trap rotation \( \Omega _{0}\gg \Omega _{c} \),
where \( \Omega _{c} \) is the critical velocity corresponding to the appearance
of the first vortex. In this case the equilibrium number of vortices \( N_{v}\gg 1 \),
and their spatial distribution is uniform.

\begin{figure}
{\par\centering \resizebox*{0.8\columnwidth}{!}{\includegraphics{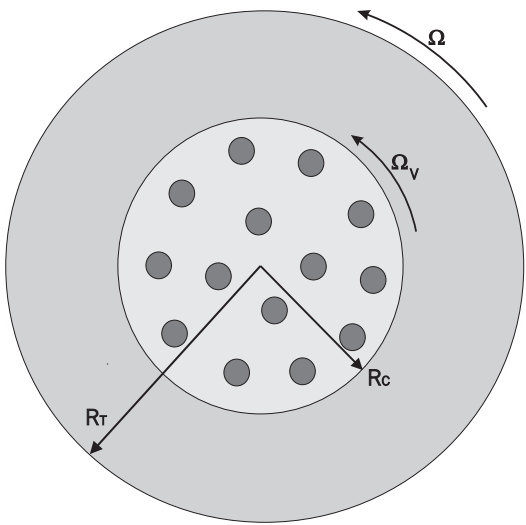}} \vspace{5mm}\par}

\caption{The schematic view of a condensate with vortices in a rotating thermal cloud.\label{fig:condensateview}}
\end{figure}

As we will see, the relaxation of vortex arrays occurs on a time scale, exceeding
the relaxation time in the thermal cloud. Hence, we may consider the evolution
of the system in quasiequilibrium approximation. This means, that at any given
moment of time, the thermal cloud is characterized by its equilibrium density
profile, corresponding to a certain temperature \( T \), and its angular velocity
of the uniform macroscopic rotation \( \Omega  \) (see Fig. \ref{fig:condensateview}).
To derive the equations of motion for the whole system, consisting of the condensate
with the vortices and the thermal cloud, we first write down the conservation
laws. Assuming that the magnetic trap does not dissipate energy, we write down
the conservation of the total energy
\begin{equation}
\label{energy:1}
\frac{d}{dt}\left( \frac{I(T)\Omega ^{2}}{2}+E(T)+E_{v}\right) =0,
\end{equation}
where \( E_{v} \) is the energy of the condensate with the vortices (see below),
\begin{equation}
\label{ET}
E(T)=3T\frac{\zeta (4)}{\zeta (3)}N(T),
\end{equation}
is the internal energy of the non-condensed particles, 
\begin{equation}
\label{IT}
I(T)=mR_{T}^{2}N(T)\frac{\zeta (4)}{\zeta (3)},
\end{equation}
is the momentum of inertia and \( R_{T}=(2T/m\omega _{\rho }^{2})^{1/2} \)
is the radial size of the thermal cloud. Here 

\begin{equation}
\label{NT}
N(T)=\frac{\zeta (3)T^{3}}{\tilde{\omega }^{3}}=\frac{T^{3}}{T_{c}^{3}}N
\end{equation}
is the number of thermal particles, \( \tilde{\omega }=(\omega _{z}\omega _{\rho }^{2})^{1/3} \),
\( T_{c}=\tilde{\omega }(N/\zeta (3))^{1/3} \) is the BEC transition temperature,
and \( N \) is the total number of particles in the trapped gas. Eqs. (\ref{ET}),(\ref{IT})
imply that \( T\gg \mu  \), where \( \mu =n_{0}g \) is the chemical potential,
\( g=4\pi \hbar ^{2}a/m \), \( a \) is the two-body scattering length, and
\( m \) is the mass of the gas particles. At lower temperatures, \( T\alt \mu  \),
the size of the thermal cloud matches the size of the condensate and, hence,
the temperature measurements become very difficult. Accordingly, in our quantitative
discussions we confine ourself to the case \( T\gg \mu  \).

In the Thomas-Fermi limit (\( \mu \gg \hbar \omega _{z,\rho } \)) the condensate
is characterized by the longitudinal \( R_{z}=(2\mu /m\omega _{\rho }^{2})^{1/2} \)
and the transverse \( R_{c}=(2\mu /m\omega _{z}^{2})^{1/2} \) sizes, so that
\( R_{c,z}\gg l_{c} \), where \( l_{c}=(\hbar ^{2}/m\mu )^{1/2} \) is the
correlation (healing) length in the condensate. The velocity of the superfluid
flow, generated by a rotating vortex grid, imitates the flow of a normal liquid
with the angular velocity \( \Omega _{v} \): \( {\bf v}_{s}=[\Omega _{v},{\bf r}] \).
The corresponding density of the vortex lines is given by \cite{LL:volIX}
\begin{equation}
\label{eqdens}
n_{v}=\frac{\Omega _{v}}{\pi \kappa },
\end{equation}
where \( \kappa =\hbar /m \) is the vortex circulation (vortices with a larger
circulation are unstable \cite{LL:volIX}). The angular velocity \( \Omega _{v} \)
must not be very large. Indeed, the maximum velocity of the superflow, which
coincides with the velocity of the superfluid on the border of the condensate
\( \sim \Omega _{v}R_{c} \) must not reach the velocity of Bogolyubov sound
\( c_{s}=\sqrt{\mu /m} \), i.e.
\begin{equation}
\label{omegamax}
\Omega _{v}\ll \frac{c_{s}}{R_{c}}\sim \omega _{\rho },
\end{equation}
since otherwise the superflow becomes unstable with respect to the spontaneous
creation of excitations already at \( T=0 \) (see Landau arguments in \cite{LL:volIX};
the nature of the rotational friction in such a situation is an interesting
question by itself and is not considered here, see \cite{volovik:rotfriction}
for a detailed discussion). Since the angular velocity \( \Omega _{v} \) is
directly related with the density of the vortex array (see Eq. (\ref{eqdens})),
the condition (\ref{omegamax}) limits the maximum number of vortices in a stable
vortex configuration confined in a harmonic trap of a given frequency:
\begin{equation}
\label{NofOmega}
N_{v}\sim \pi R_{c}^{2}\frac{\Omega _{v}}{\pi \kappa }\alt \frac{\mu }{\hbar \omega _{\rho }}\gg 1,
\end{equation}
which is a large number in the Thomas-Fermi limit.

Since the density of the condensate energy is logarithmically singular in the
vicinity of the vortex line, the energy of the vortex array, \( E_{v} \), consists
of two parts:
\begin{equation}
\label{Evortex}
E_{v}=\int d^{2}r\rho _{s}\frac{v_{s}^{2}}{2}=E_{0}+I_{v}\frac{\Omega _{v}^{2}}{2},
\end{equation}
where \( E_{0} \) is the part of the vortex energy independent on the velocities
of the vortices, and amounts to \( N_{v} \) times energy of a single vortex
at rest, \( E_{v0}=\pi mn_{0}R_{z}\log (R_{c}/l_{c}) \), see \cite{LL:volIX}.
The quantity \( I_{v}=mn_{0}R_{z}\pi R_{c}^{4}/2 \) can be called as the momentum
of inertia for the condensate with the vortices (moment of inertia of the vortex
array). The ratio of the first to the second term in the r.h.s. of Eq.(\ref{Evortex})
is \( \sim \log (R_{c}/l_{c})/N_{v}\ll 1 \). This shows that in a sufficiently
large vortex array, \( N_{v}\gg 1 \), its rest energy is negligible and the
whole internal energy of the condensate coincides with the energy of its macroscopic
rotation. 

The second equation comes from the angular momentum balance
\begin{equation}
\label{momentum:1}
\frac{d}{dt}(I(T)\Omega +M_{v})=-\alpha \Omega ,
\end{equation}
where \( M_{v} \) is the angular momentum of the vortex array and the phenomenological
coefficient \( \alpha  \) describes the interaction of the thermal particles
scattered by the trap anisotropy. In fact, the form of r.h.s. in Eq. (\ref{momentum:1})
corresponds to the dissipation of the angular momentum by a friction force proportional
to the velocity of the gas cloud. This relation between the friction force and
the gas velocity is a very general one and holds for a friction force acting
on an obstacle in a moving media both in the collisionless and hydrodynamic
limits \cite{LL:volVI}, at least for sufficiently small velocities. The choice
of the friction coefficient \( \alpha  \) depends on the details of the anisotropy
model and requires a microscopic calculation (see below). 

Since in the reference frame rotating with the superfluid the trap asymmetry
moves with a subsonic velocity (see the discussion leading to Eq.(\ref{omegamax})),
the condensate particles are not scattered by the trap and, therefore, no term
proportional to the condensate angular velocity can appear in the r.h.s. of
Eq.(\ref{momentum:1}).

Unlike in the calculation of the condensate energy (\ref{Evortex}), the expression
for the angular momentum of a vortex has no singularity and, hence, the angular
momentum of the vortex array coincides with he angular momentum of the superfluid
\begin{equation}
\label{Mvortex}
M_{v}=\int d^{2}r\rho _{s}rv_{s}=I_{v}\Omega _{v}.
\end{equation}
 Eqs. (\ref{energy:1}),(\ref{momentum:1}) are still insufficient, since they
comprise a system of two equations for the three unknown functions \( \Omega (t) \),
\( \Omega _{v}(t) \) and \( T(t) \). The formulation of the model is completed
by the addition of the equation of motion for the vortices. It follows from
the Magnus law with the friction force originating from the interaction of the
vortex lines with the thermal component. In fact it is the continuity equation
for the vortex gas moving towards the border of the condensate (see Appendix
\ref{app:dissdyn} for the derivation, which closely follows the arguments given
in \cite{fedichev:varray1}):
\begin{equation}
\label{vgrid:1}
\frac{d\Omega _{v}}{dt}+\gamma \Omega _{v}(\Omega _{v}-\Omega )=0,
\end{equation}
where \( \gamma =2\kappa \rho _{s}D/(\kappa ^{2}\rho ^{2}+D^{2}) \), \( \rho _{s}=mn_{0} \)
and \( \rho  \) are the superfluid density and the total density of the gas,
and \( D \) is the mutual friction coefficient, characterizing the interaction
of the vortex line with the excitations. 

Eqs.(\ref{energy:1}),(\ref{momentum:1}) and (\ref{vgrid:1}) constitute the
complete set of equations describing the dissipative dynamics of vortex arrays
interacting with thermal particles. The initial condition at \( t=0 \), when
the rotation of the trap stops (i.e. when the stirring beam is off), are
\begin{equation}
\label{initcond}
\Omega (0)=\Omega _{v}(0)=\Omega _{0},\; T(0)=T_{0}>\mu .
\end{equation}

The formulation of the model should be completed by ascribing a certain value
to the friction coefficient \( \alpha  \). This can be done, using the results
of the discussion of the rotational properties of trapped gas, presented in
\cite{godelin:kinetics}. Consider an ideal gas, splined up in an infinitely
long trap, characterized by the two close transverse frequencies \( \omega _{x} \)
and \( \omega _{y} \), so that \( \omega _{x,y}=\omega (1\pm \epsilon )^{1/2} \),
where \( \epsilon  \) is the parameter of anisotropy. Then, after the rotation
of the trap stops, the gas continues rotation with the angular velocity, decreasing
exponentially with the characteristic time constant given by the expression
\cite{godelin:kinetics}:
\begin{equation}
\label{tauodelin}
\lambda ={\rm Re}\frac{1}{4\tau }\left( 1-\sqrt{1-\frac{\epsilon ^{2}}{\epsilon _{c}^{2}}}\right) ,
\end{equation}
where \( \epsilon _{c}=1/4\omega \tau  \) and \( \tau  \) is the characteristic
time between the collisions in the thermal cloud. The relaxation time \( \tau  \)
can be estimated as \( \tau ^{-1}\sim n\sigma \sqrt{T/m} \), where \( n \)
is the gas density and \( \sigma  \) is the collisional cross section. To generalize
this result for the time \( \tau  \), obtained for a gas sample at \( T>T_{c} \),
for the case of a Bose condensed gas, we change the total density of the gas
to the condensate density (see the argumentation in \cite{fedichev:hydro}).

In our model, when the condensate is absent, or the interaction between the
vortices and the thermal component is negligible (this corresponds to \( \gamma \sim D=0 \)),
the vortices and the thermal excitations evolve independently. From Eq.(\ref{energy:1})
for the temperature change in the thermal cloud we have 
\begin{equation}
\label{DeltaT}
\frac{\Delta T}{T}\sim \frac{I\Omega ^{2}}{E(T)}\sim \frac{\Omega ^{2}}{\omega _{\rho }^{2}}\ll 1.
\end{equation}
The latter inequality follows from Eq.(\ref{omegamax}) and guarantees that
the temperature of the system does not change. In this approximation Eq. (\ref{momentum:1})
reads 
\begin{equation}
\label{freeeq}
\frac{d\Omega }{dt}=-\frac{\alpha }{I}\Omega ,
\end{equation}
and gives the following simple decay law for the angular velocity of the thermal
cloud 
\begin{equation}
\label{freeevol}
\Omega (t)=\Omega (0)\exp (-t/\tau _{A}),
\end{equation}
where \( \tau _{A}=I/\alpha  \) is the relaxation time related to the friction
of the moving gas and the trap anisotropy. Introducing the requirement, that
the presented model recovers the result of Guery-Odelin \cite{godelin:kinetics},
we set \( \tau _{A}^{-1}=\lambda  \) and find, that \( \alpha =I\lambda  \). 

The appearance of imaginary part in the relaxation rate (\ref{tauodelin}) for
\( \omega \tau \gg \epsilon ^{-1} \) implies an oscillatory behavior in the
onset of equilibrium of a free gas cloud (see \cite{godelin:kinetics} for the
discussion). As we shall see, the evolution of the combined system of vortices
and thermal excitations explicitly depends on the relaxation rate (\ref{tauodelin})
only in the so called ``weak anisotropy'' limit, when both the vortex array
and the non-condensed gas rotate with the same angular velocity. Since, according
to Eq. (\ref{NofOmega}), the number of vortices in the vortex array is proportional
to the angular velocity \( \Omega _{v} \), such oscillations would imply either
periodic oscillations of the vortex density (if the oscillations of \( \Omega _{v} \)
are sufficiently small), or a reversible destruction and nucleation of vortices
close to the border of the condensate. Well below \( T_{c} \), the nucleation
of vortices \cite{Donelly} is slow compared with kinetic times (\ref{tauodelin}),
and, thus, the number of vortices, as well as the angular velocity of the condensate
rotation, can only decrease in the course of dissipative dynamics. Although
the detailed description requires a more thorough discussion, we neglect the
oscillations and keep \( \alpha  \) purely real in the remaining discussion.
Collective oscillations of vortices and the thermal cloud are reviewed in \cite{Donelly}.

\section{Analysis of the dissipative dynamics.}

As discussed in the previous section, Eqs. (\ref{energy:1}),(\ref{momentum:1})
and (\ref{vgrid:1}) form a close set of equations describing the evolution
of a large (\( N_{v}\gg 1 \)) vortex array emerged in a rotating thermal cloud.
Although the equations look simple, their analytical solution requires certain
simplifications. 

First of all we qualitatively analyze the heating of the thermal cloud in the
course of dissipation of rotational energy of the gas. According to Eq. (\ref{DeltaT})
the rotational energy of the thermal cloud is always less than its internal
energy and therefore makes no contribution to the heating of the system. Thus
we can neglect it in the energy conservation law and express the temperature
of the thermal cloud from Eq. (\ref{energy:1}) as a function of angular velocity
of the condensate 
\begin{equation}
\label{TofOmegav}
T=T_{0}\left[ 1+A\left( 1-\frac{\Omega _{v}^{2}}{\Omega _{0}^{2}}\right) \right] ^{1/4}.
\end{equation}
 where \( A \) is the ratio of the rotational energy of the condensate to the
internal energy of the thermal cloud. It can be found with the help of Eqs.
(\ref{ET}) and (\ref{Mvortex})
\begin{equation}
\label{A}
A=\frac{I_{v}\Omega _{0}^{2}}{2E}\sim \left( \frac{\mu }{T_{0}}\right) ^{4}\frac{\Omega _{0}^{2}}{\omega _{\rho }^{2}}\frac{1}{(n_{0}a^{3})^{1/2}}
\end{equation}
 and can well be large. The dimensionless parameter \( A \) indicates how important
are the heating effects in the course of the dissipative dynamics. On the other
hand, it can be shown that the temperature increase \( T_{f}-T_{0}\ll T_{c} \).
This allows us to neglect the temperature dependence of \( \rho _{s} \) and
\( I_{v} \). \emph{}

Remarkably, in Eq. (\ref{momentum:1}) we can neglect also the time dependence
of the momentum of inertia of a thermal cloud. Indeed, the term containing the
time derivaty of the temperature, \( \dot{I}\Omega  \), turns to be always
small compared to \( I_{v}\dot{\Omega }_{v} \): \( \dot{I}\Omega /I\dot{\Omega }_{v}\sim (\Omega _{v}\Omega /\Omega _{0}^{2})(\omega _{\rho }/\Omega _{0})^{2}\ll 1 \).
Using this fact in Eqs. (\ref{momentum:1}) and (\ref{vgrid:1}) we obtain 

\begin{equation}
\label{cloud}
\frac{d}{dt}(\Omega -\Omega _{v})=-\frac{\alpha }{I}\left[ \eta \frac{\Omega }{\Omega _{0}}(\Omega -\Omega _{v})+\Omega \right] 
\end{equation}
where the dimensionless parameter 
\[
\eta =\frac{(I+I_{v})\gamma \Omega _{0}}{\alpha }=\frac{\tau _{R}}{\tau _{v}},\]
characterizes the relative strength of the mutual interaction between the vortices
and the thermal particles and the interaction of the thermal cloud and the trap
asymmetry. Here \( \tau ^{-1}_{v}=\gamma \Omega _{0} \) and \( \tau ^{-1}_{R}=\alpha /(I+I_{v}) \)
are the characteristic relaxation times, determined respectively by the interaction
of thermal component with the vortices and with the trap anisotropy. 

The parameter \( \eta  \) distinguishes two limiting regimes of relaxation.
First we turn to the static trap limit, when \( \tau _{R}\ll \tau _{v} \) (\( \eta \ll 1 \)).
In this case at the initial stage the friction of the thermal cloud on the trap
anisotropy is much stronger than the friction on the condensate. Therefore,
at this stage the relaxation of the thermal cloud should proceed in the same
way as in the absence of vortices, i.e. the angular velocity \( \Omega  \)
exponentially drops on the time scale \( \tau _{A}\alt \tau _{R}\ll \tau _{v} \),
while the angular velocity of a condensate remains almost unchanged, \( \dot{\Omega }_{v}\approx 0 \).
Neglecting the first term in the brackets of r.h.s of Eq. (\ref{cloud}) and
using Eq. (\ref{momentum:1}) we reduce Eq.(\ref{cloud}) to Eq. (\ref{freeeq}).
This qualitative picture remains valid until the thermal clod nearly stops and
its angular velocity reaches the very small value \( \alt \eta \Omega ^{2}_{v}/\Omega _{0}\ll \Omega _{0} \).
At this point the trap becomes effectively static and, setting \( \Omega \approx 0 \)
in Eq. (\ref{vgrid:1}), we find the equation of motion for the condensate angular
velocity

\begin{equation}
\label{stattrap}
\frac{d\Omega _{v}}{dt}=-\gamma \Omega _{v}^{2}.
\end{equation}
If the time dependence of the coefficient \( \gamma  \) can be neglected, that
either for \( A\ll 1 \), or for very small times when \( \Omega _{v}\approx \Omega _{0} \),
the solution of this equation is given by the expression
\begin{equation}
\label{omegavstatic}
\Omega _{v}(t)=\frac{\Omega _{0}}{1+t/\tau _{v}},
\end{equation}
which recovers the result previously obtained in \cite{fedichev:varray1}. 

To analyze the solutions of Eq. (\ref{stattrap}) in the strong heating limit,
\( A\gg 1 \), we note that \( \gamma \propto T \) and rewrite Eq. (\ref{stattrap})
with the help of Eq. (\ref{TofOmegav}) and the dimensionless units for the
condensate angular velocity \( \tilde{\Omega }_{v}=\Omega _{v}/\Omega _{0} \)
and the time \( \tilde{t}=t/\tau _{R}(T_{0}) \): 
\begin{equation}
\label{Ovdot_static}
\frac{d\tilde{\Omega }_{v}}{d\tilde{t}}=-(1+A(1-\tilde{\Omega }_{v}^{2}))^{1/4}\tilde{\Omega }_{v}^{2}.
\end{equation}
 Integrating it we find the exact implicit solution \( \Omega _{v}(t) \):
\[
\frac{(1+A-A\tilde{\Omega }_{v}^{2})^{3/4}}{(1+A)\tilde{\Omega }_{v}}+\frac{A\tilde{\Omega }_{v}}{2(1+A)^{5/4}}F_{\frac{1}{2};\frac{1}{4};\frac{3}{2}}\left( \frac{A\tilde{\Omega }_{v}^{2}}{1+A}\right) =\tilde{t}+C,\]
where \( F_{a,b,c}(x) \) is the hypergeometric function and the integration
constant \( C \) is to be determined from the initial condition a \( \tilde{t}=0 \).
In the considered limit of large \( A \) we find, that
\[
C=\frac{\sqrt{\pi }\Gamma (3/4)}{4\Gamma (5/4)A^{1/4}}.\]
Then, for small \( \tilde{\Omega }_{v} \), which corresponds to large times,
when the rotation of the condensate is almost stopped, the hypergeometric function
can be expanded in series of its small argument and in the lowest order in \( 1/A \)
we obtain the asymptotic expression for the condensate angular velocity
\[
\Omega _{v}=\frac{\Omega _{0}}{0.4+t/\tau _{v}(T_{f})},\]
which is valid as soon as \( t\gg \tau _{v}(T_{f}) \). This means, that the
quantity \( \tau _{v}(T_{f}) \) plays the role of the vortex array life time
and is mainly determined by the friction coefficients at the final temperature.
This fact is fairly intuitive, since the strength of the interaction of the
vortices with the thermal cloud quickly increases with the rise of the temperature
and thus most of the relaxation of the system occurs when its temperature is
close to its final value. Remarkably, the dynamics of the vortex array in a
static trap limit, that is when anisotropy is strong, turns out to be independent
from the parameters of the anisotropy.

In the opposite, rotating trap limit, when the anisotropy is weak and \( \tau _{R}\gg \tau _{v} \)
(\( \eta \gg 1 \)), the friction between the vortices and the thermal cloud
is stronger than the friction of the thermal excitation on the trap asymmetry.
As the result, the both components stick to each other due to their mutual interaction
and dissipate the angular momentum together, as a single rigid body. Neglecting
the last term in the brackets of r.h.s. of Eq. (\ref{cloud}) for the initial
conditions given by (\ref{initcond}) we find the solution \( \Omega =\Omega _{v} \).
Using this relation between the angular velocities in Eq.(\ref{momentum:1})
we arrive at
\begin{equation}
\label{rottrap}
\frac{d\Omega _{v}}{dt}=-\frac{\alpha }{I+I_{v}}\Omega _{v}.
\end{equation}
Again, when the heating of the gas sample is small, the coefficient \( \alpha  \)
is time independent, and the solution of this equation is simple:
\begin{equation}
\label{omegavrotat}
\Omega _{v}(t)=\Omega _{0}\exp (-t/\tau _{R}),
\end{equation}
i.e. in this, ``rotating trap'' approximation, the power law (\ref{omegavstatic})
changes into the simple exponential decrease of the angular velocities, with
the same time constant for both the vortex array and the thermal cloud. 

To discuss the solutions of Eq.(\ref{rottrap}) in the strong heating limit
\( A\gg 1 \) we have to make certain assumptions about the trap anisotropy.
For example, according to Eq.(\ref{tauodelin}), when the trap asymmetry is
small, \( 4\epsilon \omega \tau \ll 1 \), we have \( \alpha \propto T^{2} \).
Then Eq.(\ref{rottrap}) has the solution 
\begin{equation}
\label{hydrosol}
\Omega _{v}=\Omega _{0}\frac{2\exp (-t/\tau _{R}(T_{f}))}{1+\exp (-2t/\tau _{R}(T_{f}))}.
\end{equation}
 This means that again, the relaxation time at the finite temperature \( \tau _{R}(T_{f}) \)
plays the role of the vortex array life time. At larger times, \( t\gg \tau _{v}(T_{f}) \),
the angular velocity is small and the solution has an approximate asymptotic
form 
\[
\Omega _{v}=2\Omega _{0}\exp (-t/\tau _{R}(T_{f})).\]
Remarkably, in logarithmical approximation the solution (\ref{hydrosol}) holds
also for a situation, when the trap asymmetry is strong, \( 4\epsilon \omega \tau \gg 1 \),
and \( \alpha \propto T^{6} \) (see Eq.(\ref{tauodelin})). 

Using the presented above analysis we now can draw a qualitative picture of
dissipative dynamics. To do this, we introduce the dimensionless ratio of the
momenta of inertia of the thermal cloud and the vortex array: \( \zeta =I/I_{v}\sim (T_{0}/\mu )^{4}(n_{0}a^{3})^{1/2} \).
Since in a weakly interacting gas the gaseous parameter is small: \( (n_{0}a^{3})^{1/2}\ll 1 \),
at \( T\sim \mu  \) the condensate is heavier than the thermal cloud, and both
the rotational energy and the angular momentum are dominated by the condensate
contributions. Thus, in the rotating trap limit (\( \eta \ll 1 \)) the relaxation
time \( \tau _{R} \) is mainly determined by \( I_{v} \) and is much larger
than the spinning down time of a free thermal cloud \( \tau _{A} \) \cite{godelin:kinetics}.
In the static trap limit the role of the parameter \( \zeta  \) is more subtle.
At sufficiently large times, \( t\agt \tau _{A} \), we can neglect the term
\( I\dot{\Omega } \) in Eq. (\ref{momentum:1}), which thus transforms into
\( I_{v}\dot{\Omega }=-\alpha \Omega  \). Comparing this equation with Eq.
(\ref{stattrap}) we find that \( \Omega =\eta \Omega _{v}^{2}/\Omega _{0} \),
i.e. the thermal cloud angular velocity decays as a power law and closely follows
the relaxation of the vortex array. In the same time the heating parameter \( A \)
is related to the parameter \( \zeta  \): strong heating of the sample is only
possible when \( \zeta \gg 1 \). As the temperature grows, the number of the
condensed particles decreases and at some point the condensate may become so
small, that \( \zeta \ll 1 \). One can see, that the crossover between the
two regimes occurs at the temperature below \( T_{c} \), so the condensate
does not disappear. Later on, the heating of the system stops and the system
relaxes according to Eqs. (\ref{omegavstatic}) and (\ref{omegavrotat}) for
\( \eta \ll 1 \) and \( \eta \gg 1 \) respectively.

The presented analysis shows that the anisotropy, or any other mechanism of
angular momentum dissipation, is absolutely crucial for the onset of dissipative
dynamics. It is seen already from the fact, that when \( \tau _{R}\rightarrow \infty  \),
Eq.(\ref{omegavrotat}) gives \( \Omega _{v}(t)=\Omega (t)=\Omega _{0} \),
i.e. the rotation of the vortex grid persists forever.

Although the presented analysis is sufficient for a qualitative understanding
of the dissipative dynamics in the whole parameters range, the quantitative
description requires a numerical solution of the initial Eqs. (\ref{energy:1}),(\ref{momentum:1})
and (\ref{vgrid:1}). For the sake of completeness we provide the results of
such calculation based on the experimental parameters of \cite{vortobserv:dalibard,dalibard:varray}.
We use the following values for the model parameters: the condensate density
\( n_{0}=2\times 10^{14}cm^{-3} \), the trap frequencies \( \omega _{\rho }=2\pi \times 169s^{-1} \)
and \( \omega _{z}=2\pi \times 11.7s^{-1} \), and the temperature \( T_{0}\approx \mu =80nK \).
Then, the condensate size is \( R_{c}=4\mu m \) and, at the reported frequency
of the stirring \( \Omega _{0}=2\pi \times 135s^{-1} \), the vortex array consists
of \( N_{v}\sim 10 \) vortices. 
\begin{figure}
{\par\centering \resizebox*{0.9\columnwidth}{!}{\includegraphics{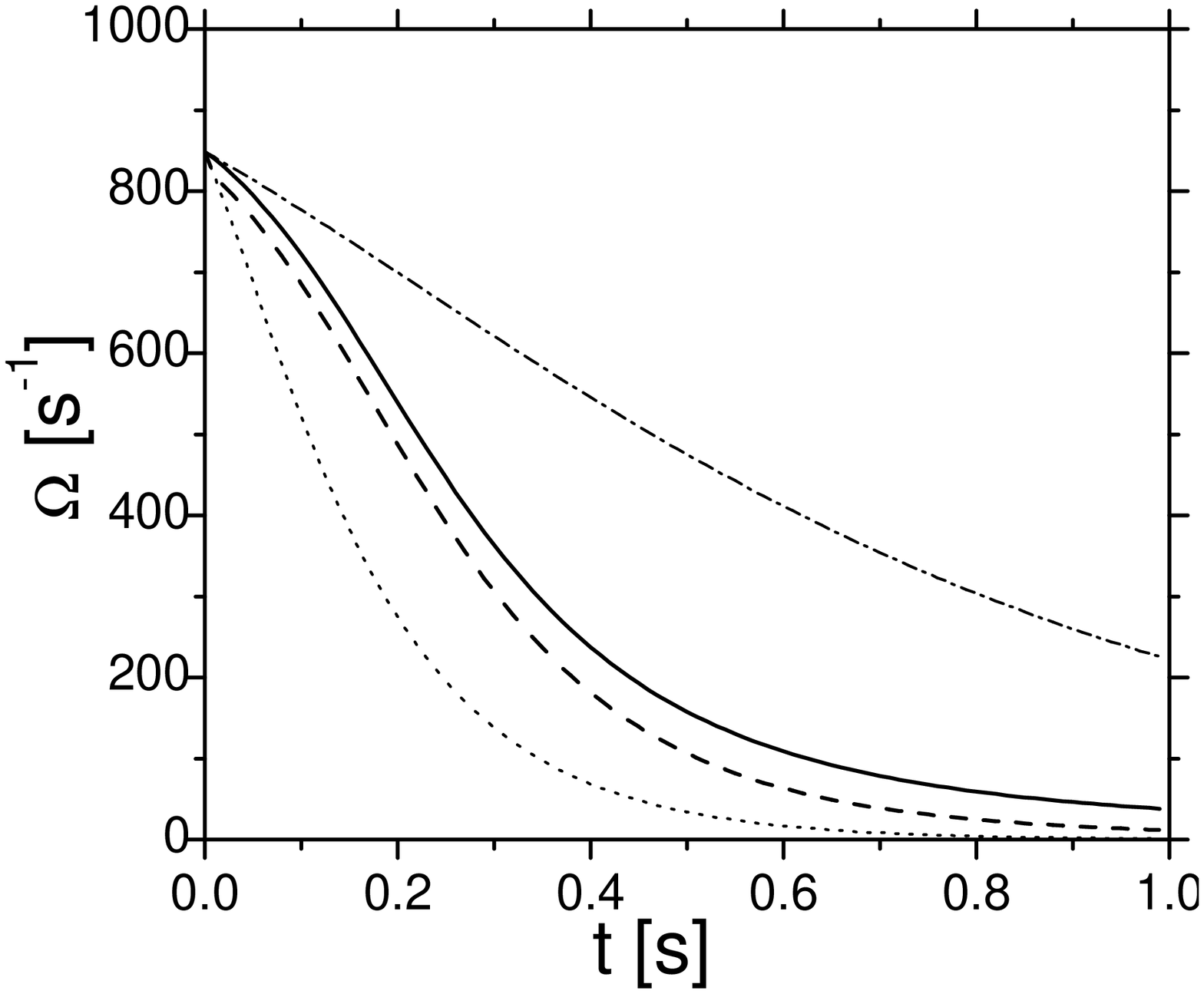}}\par }

\caption{
Solid (dashed) line represents \protect\( \Omega _{v}\protect \) 
(\protect\( \Omega \protect \))
obtained by numerical solution of Eqs. (\ref{energy:1}),(\ref{momentum:1}),
and (\ref{vgrid:1}) for the parameters of the ENS experiment.
Dotted and dash-dotted lines represent approximate solution (\ref{hydrosol})
with \protect\( \tau _{R}\protect \) calculated for the strong (\protect\( 
\epsilon \gg \epsilon _{c}\protect \))
and the weak (\protect\( \epsilon \ll \epsilon _{c}\protect \)) anisotropy
limits. \label{fig:omegas}
}
\end{figure}

\begin{figure}
{\par\centering \resizebox*{0.9\columnwidth}{!}{\includegraphics{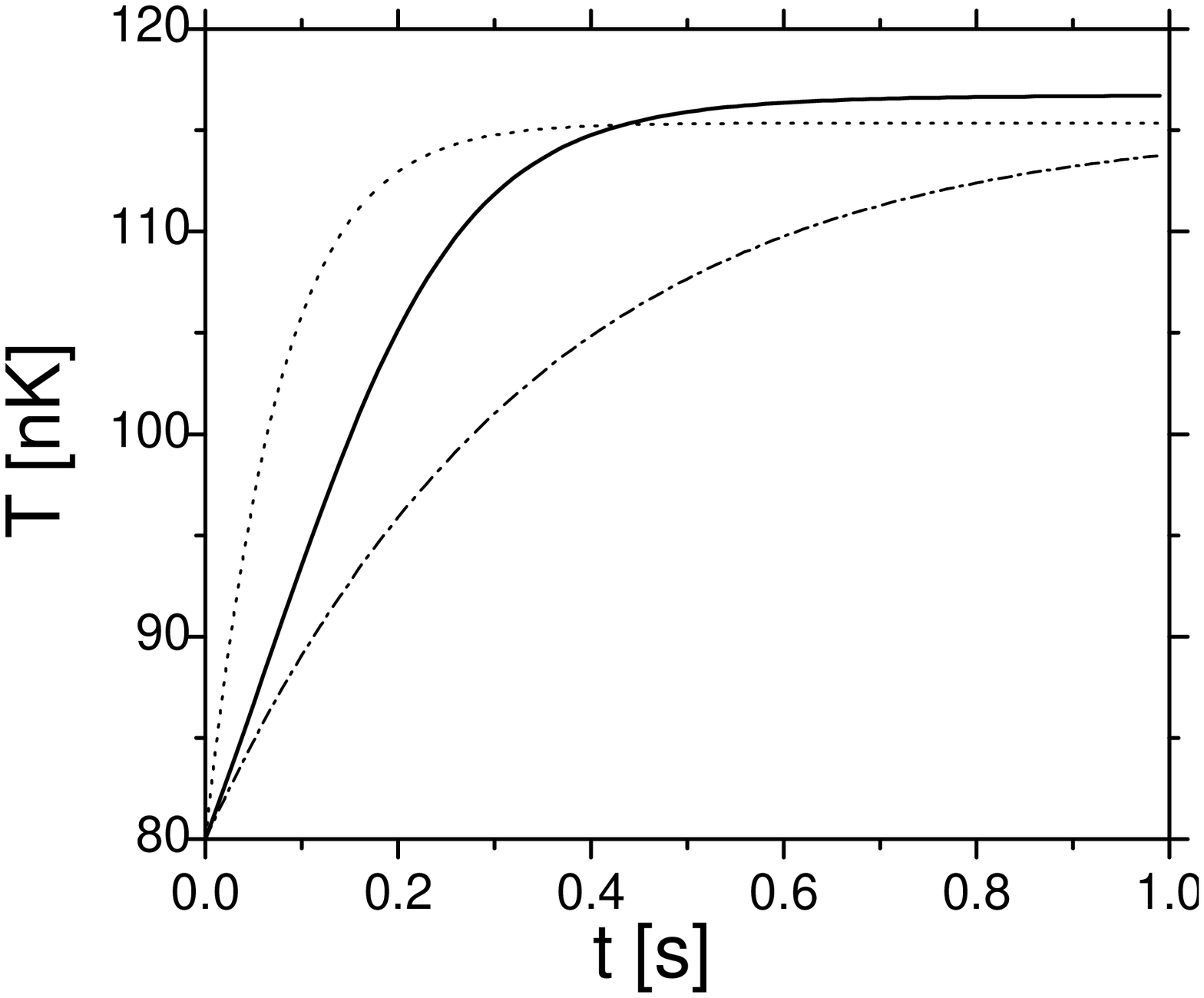}} \par}

\caption{Solid line represents temperature of the thermal cloud obtained by numerical
solution of Eqs. (\ref{energy:1}),(\ref{momentum:1}), and (\ref{vgrid:1})
corresponding to Fig. \ref{fig:omegas}. Dotted and dash-dotted lines represent
temperature obtained from Eq. (\ref{TofOmegav}) using approximate solutions
presented in Fig. \ref{fig:omegas}.\label{fig:temperature}}
\end{figure}

We provide the results of the numerical calculations for the trap anisotropy
parameter \( \epsilon =0.06 \) (see Figs. \ref{fig:omegas} and \ref{fig:temperature}).
Our model system turns out to be well in the rotating trap limit (\( \eta \sim 10 \))
and, in the full agreement with our qualitative analysis, \( \Omega \approx \Omega _{v} \)
and either of the angular velocities relaxes exponentially (see the solid and
the dashed lines on Fig. \ref{fig:omegas}). The solid line on Fig. \ref{fig:temperature}
illustrates the heating of the system in the course of the dissipative dynamics. 

The comparison of the exact numerical solution with our approximate analytical
results is a bit difficult, since this value of the anisotropy parameter is
close to \( \epsilon _{c} \) (see Eq.(\ref{tauodelin})). The dashed and the
dotted dashed lines on Figs. \ref{fig:omegas} and \ref{fig:temperature} represents
the analytical results based on Eq. (\ref{hydrosol}) with \( \tau _{v} \)
calculated in the strong \( \epsilon \gg \epsilon _{c} \) and the weak \( \epsilon \ll \epsilon _{c} \)
anisotropy limits in Eq.(\ref{tauodelin}) respectively. We note, that all of
the presented results are in a good agreement with the experimentally found
vortex array life time \( \sim 1s \).

\section{Conclusion}

In this Letter we formulate a realistic model, describing the evolution of a
large vortex array in a rotating thermal cloud interacting with the trap anisotropy.
The model allows both the simple analytical description of the most important
limiting cases as well as a very straightforward numerical analysis. 

First we show that the angular momentum dissipation mechanism, such as the trap
anisotropy is absolutely crucial for the onset of dissipative dynamics of vortices
in trapped Bose-condensed gases. We demonstrate, that if the total angular momentum
of the system is conserved, the rotation of the vortex array persists forever
and, consequently, the number of the vortices in the condensate remains constant. 

Since the flow of the condensate is a superflow, the condensate itself can not
interact with the trap anisotropy. On the contrary, the thermal excitations
are scattered by the trap and can dissipate the angular momentum. Thus, it is
just the mutual friction between the vortices and the gas particles that allows
the condensate to transfer its angular momentum to the excitations and, subsequently,
to the trap. Depending on the relative strength of interaction of the thermal
gas with vortices and with the trap anisotropy we distinguish the two regimes
of the dissipative dynamics.

First we identify the limit, when the trap anisotropy is very strong and the
rotation of the thermal cloud quickly stops. Then the vortex array interacts
with the thermal cloud at rest and subsequently looses the vortices due to the
mutual friction between the vortices and the excitations (static trap limit).
This case has been previously studied in \cite{fedichev:vortexdyn} for a single
vortex and in \cite{fedichev:varray1} for a large vortex array.

In the opposite limit, when the trap is nearly perfect, the thermal cloud stops
slowly. In this approximation the mutual friction between the vortices and the
thermal cloud makes the two components stick to each other, forming a single
rotating object with the momentum of inertia equal to the sum of the momenta
of inertia of the condensate and the thermal cloud. This composite body is usually
much heavier than the thermal cloud and hence its relaxation proceeds on a time
scale much larger than that in the Guery-Odelin scenario \cite{godelin:kinetics},
where the calculation was done above \( T_{c} \) and only the thermal cloud
was taken into account.

Another feature of a large vortex array evolution is the heating of the thermal
component caused by the energy released by the vortices escaping the condensate.
We show, that though the heating itself can be very substantial, it does not
change the qualitative picture of the dissipative dynamics. Since the friction
coefficients \( \alpha  \) and \( \gamma  \) both grow very quickly with the
temperature increase, most of the relaxation processes occurs when the temperature
is already high, i.e. when the temperature is close to \( T_{f} \). Hence,
for estimation purposes, one can use a much simpler calculation with \( T={\rm Const} \)
with all the model parameters taken at the final temperature.

\section*{Acknowledgement.}

We acknowledge fruitful discussions with G.V. Shlyapnikov, J. Dalibard and I.E.
Shvarchuck. The work was supported by Austrian Science Foundation, by INTAS,
and Russian Foundation for Basic Research (grant 99-02-18024).

\appendix

\section{Dynamics of vortex arrays.}

\label{app:dissdyn}

To derive the equation of motion for the vortices inside a rotating thermal
cloud, we closely follow the derivation presented in \cite{fedichev:varray1}.
The configuration of the superfluid is described by the superfluid velocity
\( {\bf v}_{s} \) and by the two continuous functions: the average vortex density
\( n_{v}(t,{\bf r}) \) and velocity \( {\bf v}_{v}(t,{\bf r}) \). Since the
number of vortex lines is a locally conserved quantity, the introduced functions
satisfy the continuity equation:
\begin{equation}
\label{conteq}
\frac{\partial n_{v}}{\partial t}+\frac{\partial ({\bf v}_{v}n_{v})}{\partial {\bf r}}=0.
\end{equation}

The velocity of the superfluid satisfies Maxwell-like equations
\begin{equation}
\label{Maxwell}
{\rm rot}{\bf v}_{s}=2\pi \kappa n_{v}\hat{z},\; {\rm div}{\bf v}_{s}=0,
\end{equation}
where \( \hat{z} \) is a unit vector along the axis of the cylinder \cite{LL:volIX}.
These equations are solved with the boundary condition that the normal component
of \( {\bf v}_{s} \) vanishes on border of the condensate. The vortex velocity
is related to the superfluid velocity through the Magnus law \cite{Sonin}:
\begin{equation}
\label{MagnusLaw}
\rho _{s}[{\bf v}_{v}-{\bf v}_{s},\kappa ]={\bf F},
\end{equation}
where \( \rho _{s} \) is the superfluid (mass) density and \( {\bf F} \) is
the friction force acting on the vortex line (per unit length). 

At finite temperatures, the force \( {\bf F} \) originates from the scattering
of thermal excitations from the moving vortices and consists of two terms:
\begin{equation}
\label{frictionforce}
{\bf F}=-D({\bf v}_{v}-{\bf v}_{n})+D^{\prime }[\kappa \hat{z},({\bf v}_{v}-{\bf v}_{n})],
\end{equation}
where \( {\bf v}_{n} \) is the velocity of the thermal gas at the position
of the vortex line. Here \( D \) and \( D^{\prime } \) are the longitudinal
and transverse friction coefficients respectively. Both of them are temperature
dependent and \( D^{\prime }=\rho _{n} \), where \( \rho _{n} \) is the density
of the normal component \cite{Sonin}. The friction coefficient \( D \) is
also temperature dependent and ranges from \( D\propto \kappa m^{5/2}T^{5}/\mu ^{7/2}\hbar ^{3} \)
for very small temperatures \( T\ll \mu  \) \cite{Donelly} to \( D\approx 0.1\kappa m^{5/2}T\mu ^{1/2}/\hbar ^{3} \)
\cite{fedichev:vortexdyn} for \( \mu \ll T\alt T_{c} \).

As discussed in \cite{fedichev:varray1}, the vortex density is spatially uniform:
\( n_{v}(t,{\bf r})=n_{v} \) and the superfluid velocity inside the vortex
array can be found from Eq.(\ref{Maxwell}) by using the Stokes theorem. It
has no radial component (i.e. it satisfies the necessary boundary condition)
and is given by
\begin{equation}
\label{vs}
{\bf v}_{s}({\bf r})=\pi [\kappa ,{\bf r}]n_{v}.
\end{equation}
 The solution of Eqs.(\ref{MagnusLaw}) and (\ref{frictionforce}), with \( {\bf v}_{s} \)
from Eq.(\ref{vs}) and \( {\bf v}_{n}=[\Omega ,{\bf r}] \), can be represented
in the form: \( {\bf v}_{v}=v^{(r)}{\bf \hat{r}}+v^{(\phi )}[\hat{\kappa },{\bf \hat{r}}] \),
where
\begin{equation}
\label{vvr}
v_{v}^{(r)}=(v_{s}-v_{n})\frac{\kappa \rho _{s}D}{\kappa ^{2}\rho ^{2}+D^{2}}
\end{equation}
and
\begin{equation}
\label{vvphi}
v_{v}^{(\phi )}=\frac{D^{2}v_{n}+\rho \kappa ^{2}(v_{s}\rho _{s}+D^{\prime }v_{n})}{\kappa ^{2}\rho ^{2}+D^{2}}
\end{equation}
are the radial and tangential components of the velocity field, and \( \rho =\rho _{s}+\rho _{n} \)
is the total density. Eqs. (\ref{vvr}) and (\ref{vvphi}) show, that if the
condensate rotates together with the normal component, \( v_{s}=v_{n} \), then
\( v_{v}^{(\phi )}=v_{n} \) and the radial vortex velocity is absent, i.e.
the vortex and the thermal particles move together and are in the equilibrium
in the reference frame defined by the rotating cloud.

Using Eqs.(\ref{vs}) and (\ref{vvr}), we rewrite Eq.(\ref{conteq}) in the
form
\[
\frac{dn_{v}(t)}{dt}+\gamma n_{v}(t)(\pi \kappa n_{v}(t)-\Omega )=0,\]
 where \( \gamma =2\kappa \rho _{s}D/(\kappa ^{2}\rho ^{2}+D^{2})\approx 2D/\kappa \rho  \).
Finally, expressing the vortex density through the angular velocity of the condensate
rotation (see Eq.(\ref{eqdens})), we obtain the desired equation of motion
for the vortices (\ref{vgrid:1}).

\bibliographystyle{prsty}
\bibliography{/home/fedichev/LyX/myDb}

\end{document}